\documentclass[12pt]{article}
\usepackage{graphicx}
\begin{document}

\pagestyle{empty}

\noindent
{\bf Amplitude Variations in Pulsating Yellow Supergiants}

\bigskip

\noindent
{\bf John R. Percy and Rufina Y.H. Kim\\Department of Astronomy and Astrophysics\\University of Toronto\\Toronto ON\\Canada M5S 3H4}

\bigskip

{\bf Abstract}  It was recently discovered that the amplitudes of pulsating red giants and
supergiants vary significantly on time scales of 20-30 pulsation periods.  Here, we
analyze the amplitude variability in 29 pulsating {\it yellow} supergiants (5 RVa, 4 RVb,
9 SRd, 7 long-period Cepheid, and 4 yellow hypergiant stars), using visual
observations from the AAVSO International Database, and Fourier and wavelet
analysis using the AAVSO's VSTAR package.  We find that these stars vary in
amplitude by factors of up to 10 or more (but more typically 3-5), on a mean time scale (L) of 33 $\pm$4
pulsation periods (P).  Each of the five sub-types shows this
same behavior, which is very similar to that of the pulsating red giants,
for which the median L/P was 31.
For the RVb stars, the lengths of the cycles of amplitude variability are the
same as the long secondary periods, to within the uncertainty of each.

\medskip

\noindent
{\bf 1. Introduction}

\smallskip

The amplitudes of pulsating stars are generally assumed to be
constant.  Those of multi-periodic pulsators may appear to vary
because of interference between two or more modes, though the amplitudes
of the individual modes are generally assumed to stay constant.  Polaris (Arellano Ferro 1983) and 
RU
Cam (Demers and Fernie 1966) are examples of ``unusual" Cepheids which have varied in amplitude.
The long-term, cyclic changes in the amplitudes of RR Lyrae stars --
the Blazhko effect -- are an ongoing mystery (Kolenberg 2012), period-doubling
being a viable explanation.  There are
many reports, in the literature, of Mira stars which have varied
systematically in amplitude.

Percy and Abachi (2013) recently reported on a study of the
amplitudes of almost a hundred pulsating red giants.  They found that,
in 59 single-mode and double-mode SR variables, the amplitudes of the
modes varied by factors of 2-10 on time scales of 30-45 pulsation periods, on average.
Percy and
Khatu (2014) reported on a study of 44 pulsating red {\it supergiants},
and found similar behavior: amplitude variations of a factor of up to
8 on time scales of 18 pulsation periods, on average.

In the present paper, we study the amplitudes of 29 pulsating {\it
yellow} supergiants, including 9 RV Tauri (RV) stars, 9 SRd stars, 7 long-period
Cepheids, and 4 yellow hypergiants.  RV stars show alternating deep and
shallow minima (to a greater or lesser extent).  RVa stars have constant
mean magnitude.  RVb stars vary slowly in mean magnitude; they have a ``long
secondary period".  SRd stars are semiregular yellow supergiants.  Actually,
there seems to be a smooth spectrum of behavior from RV to SRd and possibly
to long-period Population II Cepheid (Percy {\it et al.} 2003).  

Population I (Classical) Cepheids and yellow hypergiants differ from RV and
SRd stars in that they are massive, young stars, whereas the latter two classes
are old, lower-mass stars.  Classical Cepheids tend to have shorter periods
in part because the period is inversely proportional to the square root of the mass.
We did not 
analyze short-period
Cepheids because, with visual observations, it is necessary to have much denser
coverage (a large number of observations per period) in order to beat down
the observational error, which is typically 0.2-0.3 magnitude per observation.  Bright
short-period Cepheids such as $\delta$ Cep should have enough photoelectric
photometry, over time, to detect amplitude variations if they exist.  We
recommend that such a study be carried out.
We analyzed the prototype Population II Cepheid, W Vir, but the period is
short (17.27 days), and the data sparse, so the results are not very meaningful. 

The
periods of the yellow hypergiants are poorly defined, partly because the
pulsation is semi-regular at best, and partly because the light curves are
affected by the heavy mass loss and occational ``eruptions" in these stars
(e.g. Lobel {\it et al.} 2004).  Furthermore: the periods are so long that
the number of cycles of amplitude variation is very poorly-determined.

\medskip

\noindent {\bf 2. Data and Analysis}

\smallskip

We used visual observations, from the AAVSO International Database,
of the yellow supergiant variables listed in Table 1.  See ``Notes on
Individual Stars", and the last two columns in Table 1 for remarks on some of these.  Our data extend for
typically 10,000-30,000 days; not all the stars have the same length of
dataset.  Percy and Abachi (2013) discussed some of the limitations
of visual data which must be kept in mind when analyzing the observations,
and interpreting the results.  In particular: some of the stars have
pronounced seasonal gaps in the data, which can produce ``alias" periods,
and some difficulty in the wavelet analysis.

The data, extending over the range of Julian Date given in Table 1,
were analyzed with the AAVSO's VSTAR time-series analysis 
package (Benn 2013; www.aavso.org/vstar-overview), especially the Fourier
(DCDFT) analysis and wavelet (WWZ) analysis routines.  The JD range
began where the data were sufficiently dense for analysis.  The DCDFT
routine was used to determine the best period for the JD range used.
It was invariably in good agreement with the literature period; in any
case, the results are not sensitive to the exact value of period used.
For the RV stars, we used the dominant period: either the ``half" period -- 
the interval between
adjacent minima -- or the ``full" period, the interval between deep minima.
We found that, whichever of these two periods we used, the value of L/P
was the same to within the uncertainty.

For the wavelet analysis, the default values were used for the decay
time c (0.001) and time division $\Delta$t (50 days).  The results are
sensitive to the former, but not to the latter.
For the WWZ analysis:
around each of the adopted periods, we generated the amplitude versus
JD graph, and determined the range in amplitude, and the number (N)
of cycles of amplitude increase and decrease, as shown in Figures 1-10.  
N can be small and
ambiguous (see below), so it is not a precise number.

For a few stars with slow amplitude variations, we checked and confirmed the amplitude variability by using the 
DCDFT routine to determine the amplitude over sub-intervals of the range
of JD chosen.

\medskip

\noindent
{\bf 3. Results}

\smallskip

Table 1 lists the results.  It gives the name of the star, the type
of variability, the adopted period P in days, the range of JD of
the observations, the maximum and minimum amplitude, the number N of
cycles of amplitude increase and decrease, the average length L in days
of the cycles as determined from the JD range and N, the ratio L/P,
a rough measure D of the average density of the light curve relative to the
period (1 = densest, 3 = least dense), and a rough measure R of the robustness
or reliability of the amplitude versus JD curve (1 = most reliable, 3 = least reliable).
The least reliable curves have gaps, much scatter, and are generally
the ones that are least dense.  The Cepheids tended to be less reliable,
because of their shorter periods, lower density, and smaller amplitudes.  They also tend
to be less well-observed visually because observers assume that they
are best observed photoelectrically.  The yellow hypergiants are even less
reliable, because of their long periods, small amplitudes, and irregularity.
Note that the stars in Table 1 have a wide range of amplitudes, and
amplitude ranges.

The maximum and minimum amplitudes were
determined with due regard to the scatter in the amplitude versus JD curves.
The process of counting the number of cycles was somewhat subjective, but
was similar to that used by Percy and Abachi (2013) and Percy and Khatu (2014) and is therefore consistent.
Figures 1 and 2 show examples of the process for the RVa star AC Her and
the SRd star SX Her, and its uncertainty.  They also show the difference
between the amplitude versus JD curves for a shorter-period star and a
longer-period one.
The other figures show examples of the amplitude versus JD curves for 
other representative stars.

Table 2 presents summary statistics for the sub-groups of stars: Cepheids,
RVa, RVb, SRd, and hypergiant.  Note that the mean L/P is the same, within
the standard error of the mean (SEM), for all groups, and is the same as the
median L/P (31) for pulsating red giants (Percy and Abachi 2013).  The
values for the hypergiants are very uncertain, so we have listed only
approximate numbers.  The
last line (``Robust") refers to the stars whose amplitude versus JD curves
appear to be the most dependable.

\begin{figure}
\begin{center}
\includegraphics[height=5cm]{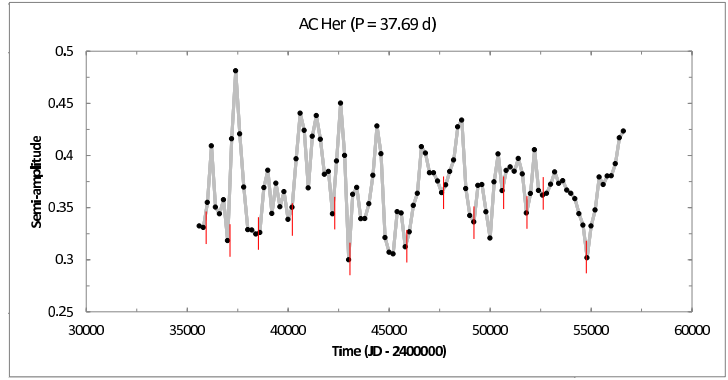}
\end{center}
\caption{Amplitude versus Julian Date for the RVa star AC Her, showing where we assume
the minima to be.  We count 12.5 cycles in a JD range of 21100 days, giving
a cycle length L of 1688 days, but the uncertainty in doing this is
apparent from the graph. The dominant pulsation period is the ``half" 
period, 37.69 days.
Compare this diagram with that for SX Her which has a longer period.}
\end{figure}

\begin{figure}
\begin{center}
\includegraphics[height=5cm]{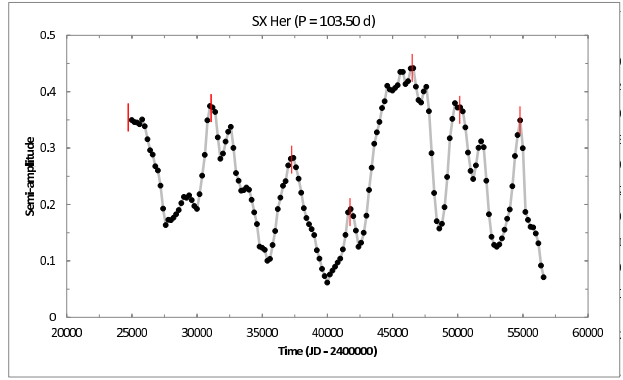}
\end{center}
\caption{Amplitude versus Julian Date for the SRd star SX Her, showing where we assume
the maxima to be.  We count 6.5 cycles in a JD range of 31631 days, giving
a cycle length L of 4866 days.  The uncertainty in doing this is
apparent from the graph, but it is less than for AC Her (Figure 1).  The pulsation period is 103.50 days.}
\end{figure}

\begin{figure}
\begin{center}
\includegraphics[height=5cm]{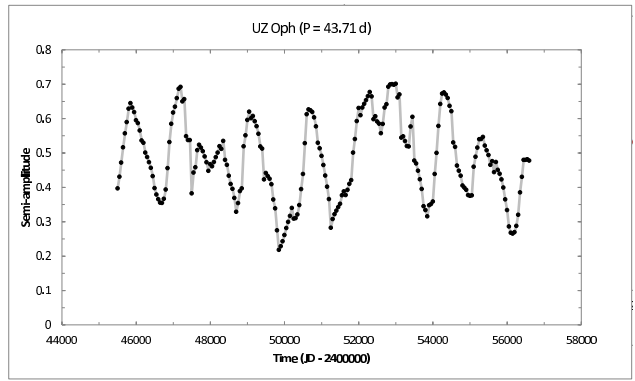}
\end{center}
\caption{Amplitude versus Julian Date for the RVa star UZ Oph.  We count
9.5 cycles.  The curve is well-defined.  The dominant pulsation period
is the ``half" period, 43.71 days.}
\end{figure}

\begin{figure}
\begin{center}
\includegraphics[height=5cm]{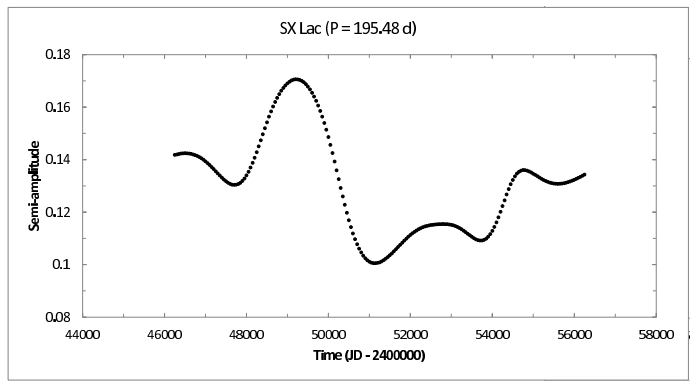}
\end{center}
\caption{Amplitude versus Julian Date for the SRd star SX Lac.  We count
3.75 cycles.  The pulsation period is 195.48 days.  Compare this graph
with e.g. the one for UZ Oph, a shorter-period star, but note that
the amplitude of SX Lac is small, and that presumably adds to the uncertainty
in determining N.}
\end{figure}

\begin{figure}
\begin{center}
\includegraphics[height=5cm]{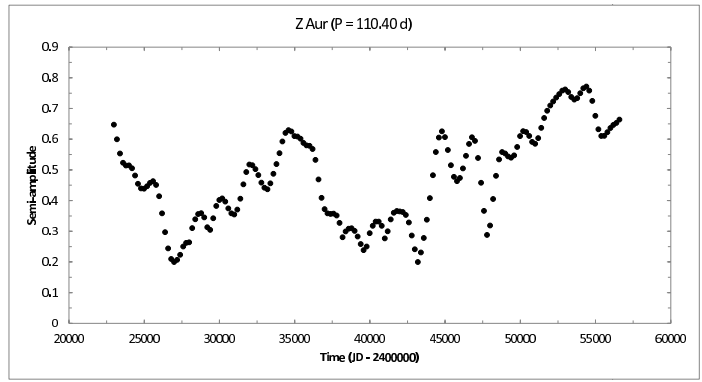}
\end{center}
\caption{Amplitude versus Julian Date for the SRd star Z Aur. We count
15 cycles.  This star is
unusual in that its dominant period switches between about 112 and 135 days.
The amplitude tends to decrease before period switches, which occured around
JD 2429000 and 2448000.}
\end{figure}

\begin{figure}
\begin{center}
\includegraphics[height=5cm]{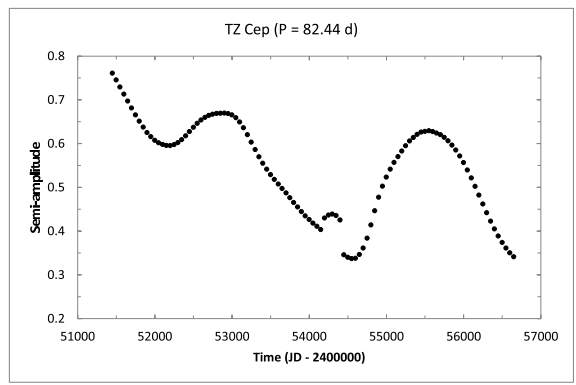}
\end{center}
\caption{Amplitude versus Julian Date for the SRd star TZ Cep.  We count 2.2
cycles.  The pulsation
period is 82.44 days.}
\end{figure}

\begin{figure}
\begin{center}
\includegraphics[height=5cm]{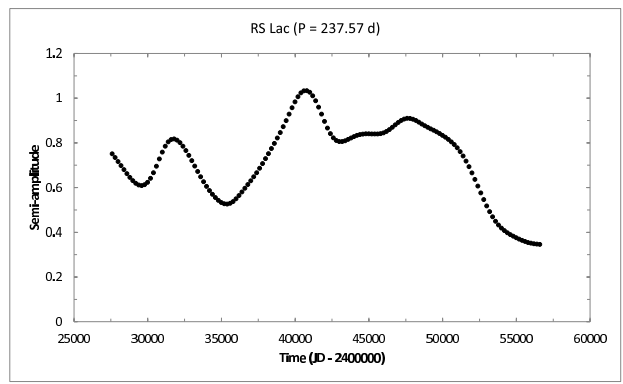}
\end{center}
\caption{Amplitude versus Julian Date for the SRd star RS Lac.  We count
3.5 cycles.  The pulsation
period is 237.57 days.}
\end{figure}

\begin{figure}
\begin{center}
\includegraphics[height=5cm]{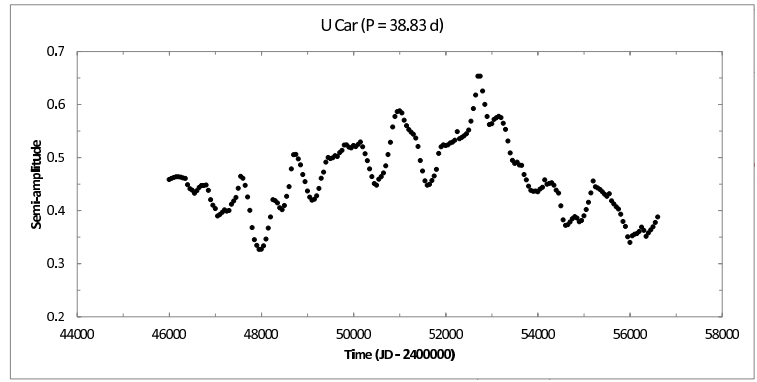}
\end{center}
\caption{Amplitude versus Julian Date for the Cepheid U Car.  We count 9 cycles.
There is a slow change in amplitude, as well as the rapid ones.  The
pulsation period is 38.83 days.  The rapid changes in amplitude are relatively
small, suggesting that the mechanism which causes them is not dominant in this
star.}
\end{figure}

\begin{figure}
\begin{center}
\includegraphics[height=5cm]{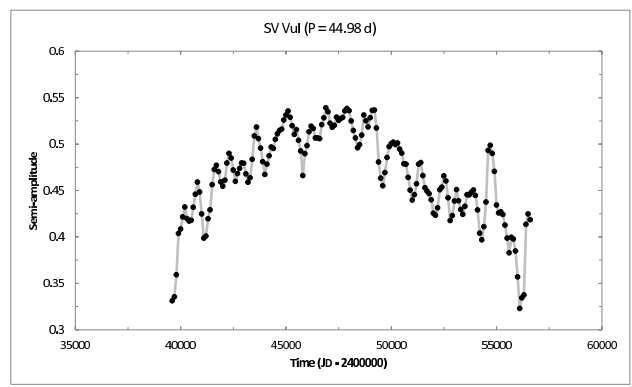}
\end{center}
\caption{Amplitude versus Julian Date for the Cepheid SV Vul.  We count 13.5
cycles.  These are small and rapid, and therefore not well-defined by our
limited visual observations.  There is also a slow change in amplitude.  The
pulsation period is 44.98 days.  The rapid changes in amplitude are relatively
small, suggesting that the mechanism which causes them is not dominant in this
star.}
\end{figure}

\begin{figure}
\begin{center}
\includegraphics[height=5cm]{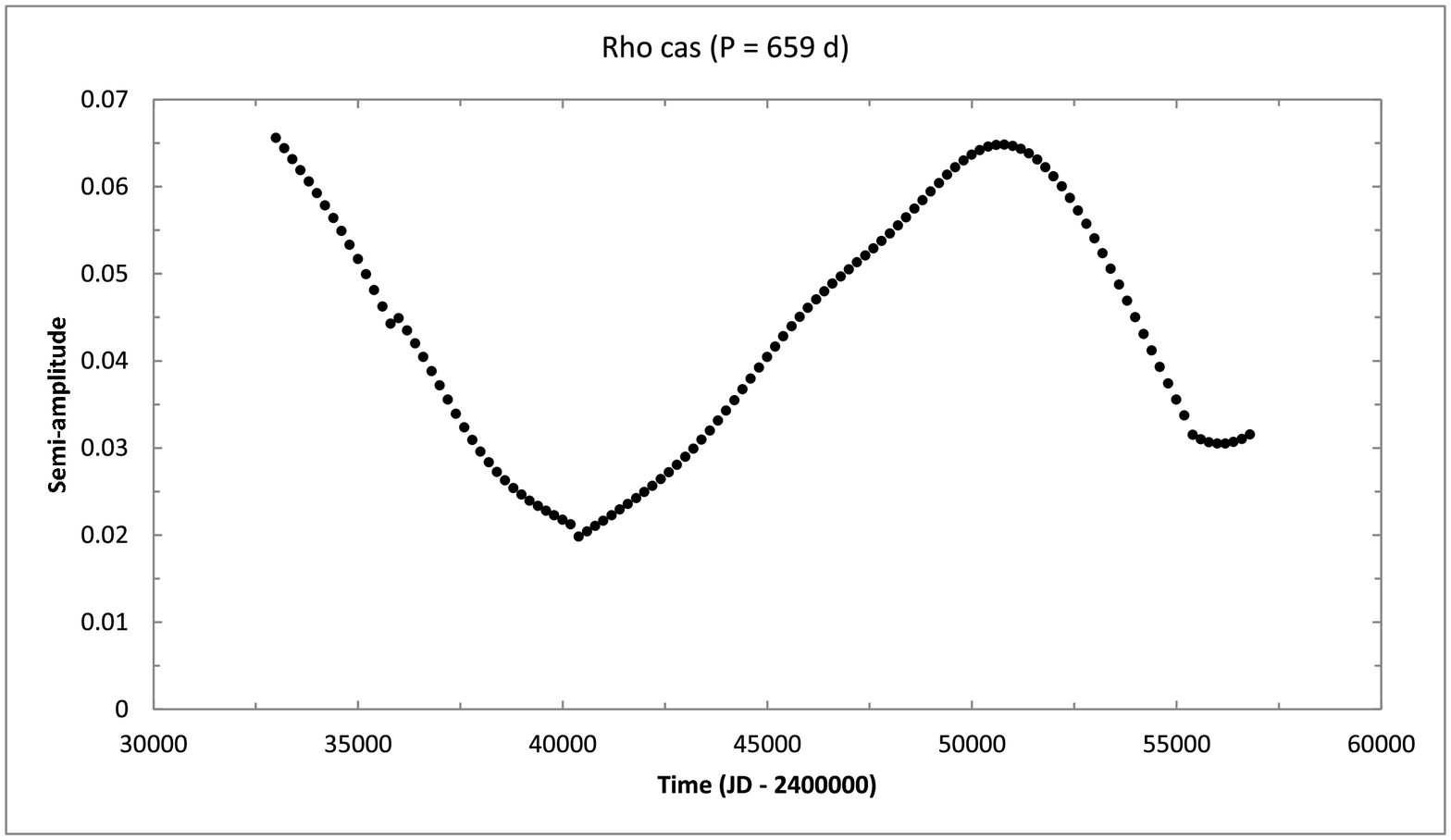}
\end{center}
\caption{Amplitude versus Julian Date for the yellow hypergiant $\rho$ Cas.
We count 1.35 cycles, though this is obviously very uncertain -- even more
so for the other yellow hypergiants. The adopted pulsation period is 659 days.}
\end{figure}

\medskip

\noindent
{\bf 3.1 Notes on Individual Stars}

\smallskip

These notes are given in the same order as the stars are listed in Table 1.
See also the last two columns in Table 1 for information about the denseness
of the light curves, and the robustness of the amplitude versus JD curves.

\smallskip

{\it SU Gem:}  The seasonal gaps are very conspicuous.

{\it AC Her:}  The increase in amplitude since JD 2455000 is confirmed by the
AAVSO photoelectric photometry.

{\it Z Aur:}  This star shows periods of 112 and 135 days, and switches
between them (Lacy 1973).  There is some evidence that the amplitude of pulsation
decreases before a switch takes place.

{\it TZ Cep:}  The data are sparse since JD 2454200.

{\it DE Her:}  This star shows 1.5 cycles of a long secondary period, but is
classified as SRd rather than RVb.

{\it UU Her:}  This star switches between periods of 45-6 and 72 days
(Zsoldos and Sasselov, 1992).

{\it RS Lac:}  The amplitude variation is apparent from the light curve.

{\it SX Lac:}  The data are initially sparse.

{\it S Vul:}  The middle of the dataset is sparse.

{\it V509 Cas:}  Later in the visual dataset, the dominant period is 259 days.
The photoelectric V data, however, show periods between 350 and 500 days.

{\it V1302 Aql:}  The period is suspeciously close to one year.  Furthermore:
the data are sparse and the amplitude is small.

\medskip

\noindent
{\bf 4. Discussion}

\smallskip

We have found that almost all of the pulsating yellow supergiants that we
have studied vary in pulsation amplitude by a factor of up to 10 on a time
scale of about 33$\pm$ pulsation periods.  The behavior is similar in each of
the subtypes of variables, and that behavior is similar to that of
pulsating red giants (Percy and Abachi 2013).  In particular: the RV Tauri
variables showed similar L/P to the other types, whether the half-period
or the full period was dominant. 
These results were pleasantly
surprising to us, as we had no {\it a priori} reason to think that these
stars would show amplitude variations or, if so, that these would be
similar to those in red giants and supergiants.  In some cases, however,
the amplitude variation in these stars is visible in the light
curve.

We note that, for the RVb stars, the lengths of the cycles of amplitude
variability are the same as the lengths of the long secondary periods,
within the uncertainties of each.  For SU Gem, L = 683, LSP = 682; for
IW Car: L = 1326, LSP = 1430; for DF Cyg: L = 780, LSP = 784; for AI Sco:
L = 977, LSP = 975, the units being days in each case.  Percy (1993) noted
that, during the long secondary minima in the RVb star U Mon, the pulsation
amplitude was low.  This coincidence between L and LSP may help to elucidate
the cause of both the RVb phenomenon, and the amplitude variation.

Percy and Abachi (2013)
proposed two possible explanations for the amplitude variation in
pulsating red giants: (i) the
rotation of a star with large inhomogeneities in its photosphere; and (ii)
stochastic excitation and decay of pulsations, driven by convection (this
possibility was suggested to us by Professor Tim Bedding).  Red giants and
supergiants are highly convective, and there is evidence (e.g. Kiss 
{\it et al.} 2006, Xiong and Deng 2007) that the convection interacts with the pulsation.  Cepheid pulsations
are excited by the kappa (opacity) mechanism; hydrodynamic models (e.g. Stobie 1969)
show that the pulsation amplitude grows until the pulsational energy
generation is balanced by dissipation.  As for the amplitude {\it variations}: since
yellow supergiants are not expected to show large inhomogeneities in their
photospheres, stochastic excitation and decay is the more likely explanation.
All of the stars in our sample are cooler than the sun (their (B-V)s range
from +0.9 to +1.8) so they all have significant external convection zones.  We note
that, in the long-period Cepheids U Car and SV Vul (Figures 8 and 9), the
amplitude fluctuations are relatively small, but there are also slow
changes in amplitude as well as the small, rapid ones.
In addition to the possibility of stochastic excitation, these stars are
subject to possible non-linear effects such as period doubling
and chaos (Buchler and Kovacs 1987, Fokin 1994, Buchler {\it et al.} 1996,
 Buchler {\it et al.} 2004).

Amplitude variations complicate the study of these stars in the sense that,
to compare photometric behavior with other types of behavior -- spectroscopic,
for instance -- the observations must be made within a few pulsation periods
of each other.  The AAVSO provides an important service by monitoring many
of these stars.

\medskip

\noindent
{\bf 5. Conclusions}

\smallskip

We have studied the amplitude variation in 29 pulsating yellow supergiants
of several types: RV Tauri stars (RVa and RVb), SRd stars, long-period
Cepheids, and hypergiants.  In each case, we find amplitude variations
of a factor of up to 10 (but more typically 3-5) on a time scale of 33
pulsation periods.  The behavior is similar for each type of star, and is
similar to that found by Percy and Abachi (2013) in pulsating red giants.

\medskip

\noindent
{\bf Acknowledgements}

\smallskip

We thank the hundreds of AAVSO observers who made the observations which were
used in this project, and we thank the AAVSO staff for processing and
archiving the measurements, and making them publicly available.  We also thank the team which developed the
VSTAR package, and made it user-friendly and publicly available.  We
thank the University of Toronto Work-Study Program for financial support.
This project
made use of the SIMBAD database, which is operated by CDS,
Strasbourg, France.

\medskip

\noindent
{\bf References}

\smallskip

\noindent
Arellano Ferro, A., 1983, {\it Astrophys. J.}, {\bf 274}, 755.

\smallskip

\noindent
Benn, D. 2013, VSTAR data analysis software (http://www.aavso.org/node/803).

\smallskip

\noindent
Buchler, J.R., and Kov\'{a}cs, G., 1987, {\it Astrophys. J.}, {\bf 320}, 57.

\smallskip

\noindent
Buchler, J.R., Koll\'{a}th, Z., Serre, T., and Mattei, J.A., 1996, {\it Astrophys. J.}, {\bf 462}, 489.

\smallskip

\noindent
Buchler, J.R., Koll\'{a}th, Z., and Cadmus, R.R. Jr., 2004, {\it Astrophys. J.}, {\bf 613}, 532.

\smallskip

\noindent
Demers, S., and Fernie, J.D., 1966, {\it Astrophys. J.}, {\bf 144}, 440.

\smallskip

\noindent
Fokin, A.B., 1994, {\it Astron. Astrophys.}, {\bf 292}, 133.

\smallskip

\noindent
Kiss, L.L., Szab\'{o}, G.M. and Bedding, T.R. 2006, {\it Mon. Not. Roy. Astron. Soc.}, {\bf 372}, 1721.

\smallskip

\noindent
Kolenberg, K., 2012, {\it JAAVSO}, {\bf 40}, 481.

\smallskip

\noindent
Lacy, C.H., 1973, {\it Astron. J.}, {\bf 78}, 90.

\smallskip

\noindent
Lobel, A. {\it et al.}, 2004, in {\it Stars as Suns: Activity, Evolution, and
Planets}, IAU Symposium 219, Astronomical Society of the Pacific, San Francisco,
CA, 903.

\smallskip

\noindent
Percy, J.R. 1993, in {\it Non-Linear Phenomena in Stellar Variability}, ed.
M. Takeuti and J.-R. Buchler, Kluwer Academic Publishers, Dordrecht, The
Netherlands, 123.

\smallskip

\noindent
Percy, J.R., Hosick, J., and Leigh, N.W.C., 2003, {\it Publ. Astron. Soc. Pacific}, {\bf 115}, 59.
 
\smallskip

\noindent
Percy, J.R. and Abachi, R., 2013, {\it JAAVSO}, {\bf 41}, 193. 

\smallskip

\noindent
Percy, J.R. and Khatu, V., 2014, {\it JAAVSO}, in press.

\smallskip

\noindent
Stobie, R.S., 1969, {\it Mon. Not. Roy. Astron. Soc.}, {\bf 144}, 461.

\smallskip

\noindent
Xiong, D.R., and Deng, L., 2007, {\it Mon. Not. Roy. Astron. Soc.}, {\bf 378}, 1270.

\smallskip

\noindent
Zsoldos, E., and Sasselov, D.D., 1992, {\it Astron. Astrophys.}, {\bf 256}, 107.

\small

\begin{table}\small
\caption{Amplitude Variability of Pulsating Yellow Supergiants.} 
\begin{tabular}{rrrrrrrrr}
% \multicolumn{9}{c}{Table 1. Amplitude Variability of Pulsating Red Supergiants.} 
\hline
Star & Type & P(d) & JD Range & A Range & N & L/P & D & R \\
\hline
AG Aur & SRd & 96.13 & 2441000-2456600 & 0.15-0.78 & 11 & 14.8 & 2 & 2 \\
AV Cyg & SRd & 87.97 & 2429012-2456630 & 0.11-0.57 & 9 & 34.9 & 1 & 1 \\
SU Gem & RVb & 24.98 & 2446000-2456250 & 0.00-1.35 & 15 & 27.4 & 3 & 3 \\
AC Her & RVa & 37.69 & 2435500-2456600 & 0.28-0.49 & 12.5 & 44.8 & 1 & 2 \\
SX Her & SRd & 103.50 & 2425000-2456631 & 0.08-0.44 & 6.5 & 47.0 & 1 & 1 \\
TT Oph & RVa & 30.51 & 2427946-2456615 & 0.25-0.76 & 39 & 24.1 & 1 & 2 \\
UZ Oph & RVa & 43.71 & 2445500-2456626 & 0.20-0.70 & 9.5 & 26.8 & 1 & 1 \\
TX Per & RVa & 76.38 & 2427964-2456654 & 0.12-0.75 & 7.5 & 50.1 & 2 & 1 \\
V Vul & RVa & 76.31 & 2446000-2456649 & 0.20-0.35 & 6.5 & 21.5 & 1 & 1 \\
IW Car & RVb & 71.96 & 2446037-2456646 & 0.05-0.24 & 8 & 18.4 & 1 & 2 \\
DF Cyg & RVb & 24.91 & 2441000-2456600 & 0.20-0.86 & 20 & 31.3 & 2 & 1 \\
AI Sco & RVb & 35.76 & 2445000-2455750 & 0.15-1.10 & 11 & 27.3 & 2 & 2 \\
Z Aur & SR/SRd & 110.40 & 2423000-2456651 & 0.20-0.76 & 15 & 20.3 & 1 & 1 \\
TZ Cep & SRd & 82.44 & 2451426-2456635 & 0.33-0.76 & 2.2 & 28.7 & 2 & 1 \\
DE Her & SRd & 173.10 & 2442000-2456622 & 0.13-0.64 & 1.5 & 56.3 & 1 & 1 \\
UU Her & SRd & 44.93 & 2432000-2456622 & 0.02-0.18 & 15 & 36.5 & 2 & 2 \\
RS Lac & SRd & 237.57 & 2427592-2456635 & 0.35-1.02 & 3.5 & 34.9 & 1 & 1 \\
SX Lac & SRd & 195.48 & 2446000-2456282 & 0.10-0.17 & 3.75 & 14.0 & 2 & 1 \\
W Vir & CWA & 17.27 & 2425000-2456458 & 0.16-1.29 & 41 & 16.3 & 2 & 3 \\
T Mon & DCEP & 27.03 & 2441500-2456400 & 0.28-0.53 & 15 & 36.7 & 1 & 2 \\
X Pup & DCEP & 25.96 & 2434287-2456416 & 0.15-1.04 & 15 & 56.8 & 3 & 3 \\
L Car & DCEP & 35.54 & 2426096-2456609 & 0.22-0.42 & 12 & 71.5 & 1 & 3 \\
U Car & DCEP & 38.83 & 2443617-2456627 & 0.33-0.63 & 9 & 37.2 & 1 & 2 \\
S Vul & DCEP & 68.30 & 2439626-2456558 & 0.10-1.20 & 11 & 22.5 & 3 & 2 \\
SV Vul & DCEP & 44.98 & 2439622-2456637 & 0.33-0.54 & 13.5 & 28.0 & 1 & 1 \\
$\rho$ Cas & YHG & 659 & 2433000-2456723 & 0.02-0.07 & 1.35 & 26.7 & 1 & 1 \\
V509 Cas & YHG & 878 & 2446923-2456719 & 0.03-0.06 & 0.25 & 44.6 & 2 & 1 \\
V766 Cen & YHG & 871 & 2446933-2456709 & 0.09-0.15 & 0.40 & 28.1 & 1 & 1 \\
V1302 Aql & YHG & 359 & 2442542-2455368 & 0.02-0.09 & 1.5 & 23.8 & 2 & 1 \\ 
\hline
\end{tabular}
\end{table}

\small

\begin{table}\small
\caption{Summary Statistics: Amplitude Variability of Pulsating Yellow Supergiants.} 
\begin{tabular}{rrrrr}
% \multicolumn{9}{c}{Table 1. Amplitude Variability of Pulsating Red Supergiants.} 
\hline
Type & Mean P (SD) & Mean L/P (SEM) & Mean D  & Mean R  \\
\hline
Cepheids & 41.09 (14.45) & 38.46 (7.38) & 1.71 & 2.29 \\
RVa & 52.92 (21.89) & 33.45 (5.83) & 1.20  & 1.40  \\
RVb & 39.40 (22.30) & 26.11 (2.72) & 2.00  & 2.00  \\
SRd & 125.72 (2.30) & 31.94 (4.75) & 1.44  & 1.22  \\
Hypergiants & 700: & 31: & 1.5 & 1 \\
Robust & 104.73 (65.12) & 32.82 (3.67) & 1.33 & 1.00  \\
\hline
\end{tabular}
\end{table}

\end{document}